# Mg decorated Boron doped Graphene for Hydrogen Storage

Baliram Lone[1]*

[1] Nanomaterials Research Laboratory, Department of Physics, Vinayakrao Patil Mahavidyalaya Vaijapur, Dist. Aurangabad, Maharashtra 423701, India

*corresponding author E-mail:baliram.lone@aggiemail.usu.edu

**Abstract:**

First principles based DFT calculations performed to insight structural and electronic properties of Boron doped Magnesium atom decorated graphene sheet for application of hydrogen storage. The four $H_2$ molecules stably binds magnesium atom with Boron doped graphene sheet. The average binding energy extracted in the range-0.566 to -0.687 eV/H2.Partial density of states of complex system shows s and d orbitals of $H_2$ molecule and Mg atom at -0.1eV overlaps of main peaks indicates strong hybridizing and binding of s and d orbitals of $H_2$ and Mg atom respectively. The gravimetric capacity of studied complex system reaching approximately 8.26 wt% hydrogen. HOMO & LUMO study shows stability of complex system.DOS investigation reveals the electronic density of states of complex system.

## 1. Introduction

Now a days there is rapid increase of energy demand among people's daily life. To cater the need of daily consumption of fossil fuel, research on new environment-friendly energy sources and their practical applications has attracted increasing attention in the past decades [1-4]. Ensuring clean and efficient energy sources is one of the biggest challenges that we face in the 21st century.

With high energy conversion efficiency, zero pollutant emission, clean-burning product, rich in energy per unit mass, and most potentially abundant source, hydrogen energy is considered as the most clean and promising alternative energy source in the future [5].

To eliminate or remove $CO_2$ emission at end user, the use of dihydrogen ($H_2$) as energy carrier is environmentally friendly method. To realize the Hydrogen Economy [6-7], the prime goal is investigating of safe, efficient and effective stores for $H_2$ gas, and replace current technologies based around the compression of $H_2$ as a liquid or as a gas using cryogenic. Recent, theoretical studies explained the poor capacity of $H_2$ adsorption on monolayer graphene surface. The theoretical results indicate, the adsorption energy ($E_{ad}$) of $H_2$ on graphene sheets is approximately 5 kJ/mol [1], which is far from the recommended $E_{ad}$ (20-40 kJ/mol or 0.2-0.6 eV/Hydrogen molecule) for use of practical applications [7]. Therefore, enhanced adsorption energy of hydrogen on graphene could improve its uptake at room temperature [1]. Furthermore, theoretical results validates that a key factor leading to low $H_2$ adsorption capacity of pristine graphene is weak binding between graphene sheets and $H_2$ atoms under ambient conditions.

The decoration of graphene (Gr) sheets is considered as one of the promising methods for hydrogen uptake improvement at ambient temperature [8,9]. Transition metals such as palladium (Pd) [10,11], calcium (Ca) [12,14], vanadium (V) [15], titanium (Ti) [16], nickel (Ni) [17], iron (Fe) [18] and Aluminium [19] can be useful for this purpose. Among these transition metals, palladium provides the best conditions for hydrogen storage due to high catalytic activity [8] and high affinity for hydrogen sorption [20]. Pd is five times cheaper than Pt and more electrochemically stable than other transition metal atoms such as Fe, Co and Ni [21].

To cater the demand of energy in vehicular system, one of the suitable way for production of renewable source such as hydrogen storage which is air pollution free promising candidate. We explored to achieve possibility of hydrogen storage for onboard application.

In this study, for the first time a novel structure of graphene sheets as Mg-doped graphene sheets with boron in the presence of Mg atom was proposed to improve the hydrogen storage. This serves as the main novelty of this study. During hydrogen adsorption, boron can increase the adsorption of dissociated hydrogen atoms; their simultaneous application could be beneficial. Boron doping is a feasible and practical method to alter the binding structure and enhance the adsorption performance of hydrogen.

## 2. Methods

We considered the model [23] and further quantum chemical calculations were applied by using density functional theory. We applied Vienna ab initio simulation code (VASP) [24-25] within PAW method. The exchange and correlation within local density approximation have been selected for our studied system. To interpret all the characteristics of molecular interactions no density functional theory describes it accurately, particularly Van der Waals (VdW) interactions [26-28]. The LDA method employed to study the physisorption energies of $H_2$ on CNT and graphene are in good agreement with experimental values [29-30].

However, the overestimate of the binding energy by LDA is compensated by the ignored van der Waals interactions [31-32]. The electron wave functions are expanded by plane waves with a kinetic energy cutoff of 450 eV to attain the required convergence. All of the self-consistent loops are iterated until the total energy difference of the systems between the adjacent iterating steps is less than $10^{-7}$ eV. The Brillouin zone is sampled by $5 \times 5 \times 1$ mesh points in k-space based on Monkhorst-Pack scheme [33-34]. The effective range of the kinetic energy cutoff and the validity of the mesh density used in this calculation are determined by a convergence test using the theoretically estimated lattice constants of the pristine graphene, 2.46 Å. To avoid the interactions of adjacent slabs, the vacuum space of 20 Å is introduced for $4 \times 4$ supercell which contains 32 carbon atoms.

## 3. Results and discussion

### 3.1 Characterization of Magnesium (Mg)

We considered that Mg atoms distributed uniformly on graphene (Gr) sheet, However, the phenomenon of adsorption behavior of Mg on pristine graphene sheet investigated as mentioned in below equation. The adsorption energy ($E_{ad}$) of Mg on graphene (Gr) can be calculated as:

$$E_{ad-Mg} = E_{Mg/Gr} - E_{Mg} - E_{Gr} \tag{1}$$

Where $E_{Mg/Gr}$, $E_{Mg}$, and $E_{Gr}$ are evaluated total energies of the complex Mg-decorated graphene sheet, isolated Mg atom and pristine graphene (Gr) sheet. The favorable adsorption site for magnesium atom on surface of Gr sheet have centre of a hexagonal ring. However, using above formula 1, the adsorption energy ($E_{ad}$) of Mg on graphene ($E_{ad\text{-}Mg}$) is calculated and it is reported -2.88 eV.

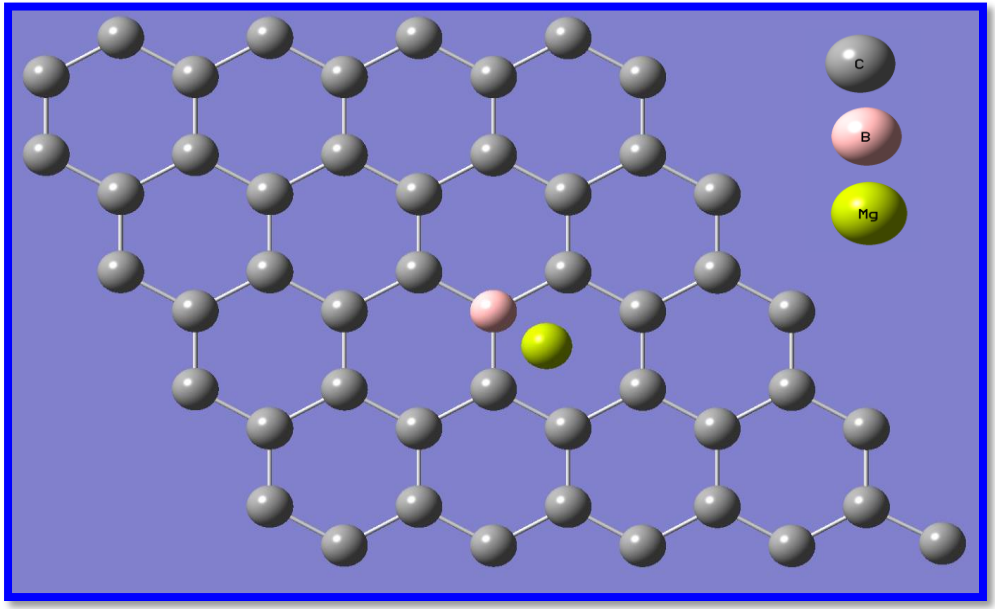

**Figure 1.** The Mg decorated stable graphene sheet with Boron doping atom

However, the cohesive energy of complex system in solid phase of Magnesium found -4.26 eV/atom is greater than evaluated adsorption energy of Magnesium atom on pristine graphene sheet.

It implies that the magnesium atoms forms cluster. We introduce Boron atom to 4x4x1 graphene supercell, to achieve stable and uniform decoration of magnesium atoms on Gr sheet. The mechanism of adsorption energy ($E_{ad}$) of Mg on Boron doped graphene sheet is evaluated by

$$E_{ad-Mg} = E_{Mg/B/Gr} - E_{Mg} - E_{B/Gr} \qquad (2)$$

Where $E_{Mg/B/Gr}$, $E_{Mg}$ and $E_{B/Gr}$ is total energies of complex system of the Mg decorated B doped graphene, isolated Mg atom and B doped graphene sheet respectively.

Under study, system is fully relaxed to achieve stable configuration stable configuration. The Mg atom tends to stay top site near the boron (B) atom as shown in figure 1. The distance between Mg and B atom is found 1.154 Å. The distance C-C 1.523Å whereas the distance between Mg and C atom is 1.334 Å.

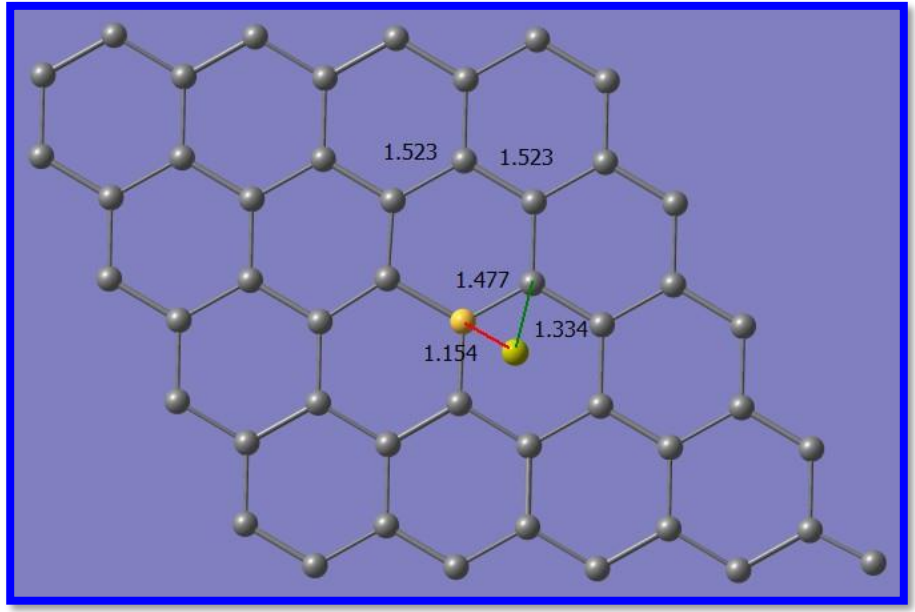

**Figure 2** .The bond distance between Mg, B & C-C in Å

The investigated simulation results reveals the adsorption energy of Mg atom on Boron doped Gr sheet is -4.48 eV/atom is greater than bulk or complex solid phase magnesium cohesive -4.26 eV/atom.However,The formation of cluster of Mg atoms on Gr sheet can be prevented successfully by Boron doping process.

### 3.2 Adsorption of hydrogen on complex system (Mg/B/Gr)

The complex system of Mg decorated boron doped graphene sheet investigated for the on board vehicular application of hydrogen storage. The adsorption of hydrogen ($H_2$) molecules on magnesium decorated graphene sheet summarized as:

After full relaxation optimized configurations of all $H_2$ molecules demonstrates in Figures 3 a-d (Top view).Figures 4a-d(side view) reveals the $H_2$ molecules prefers takes sites near to Mg atoms as well and vertical distance between first molecule and graphene sheet reported is 2.50 The other molecules exhibit same distance with the graphene sheet. The average bond length $d_{Avg}$=H-H is 0.925 Å which is elongated as shown in figure 3a, in gas phase bond length is 0.74 Å of the $H_2$ molecules.

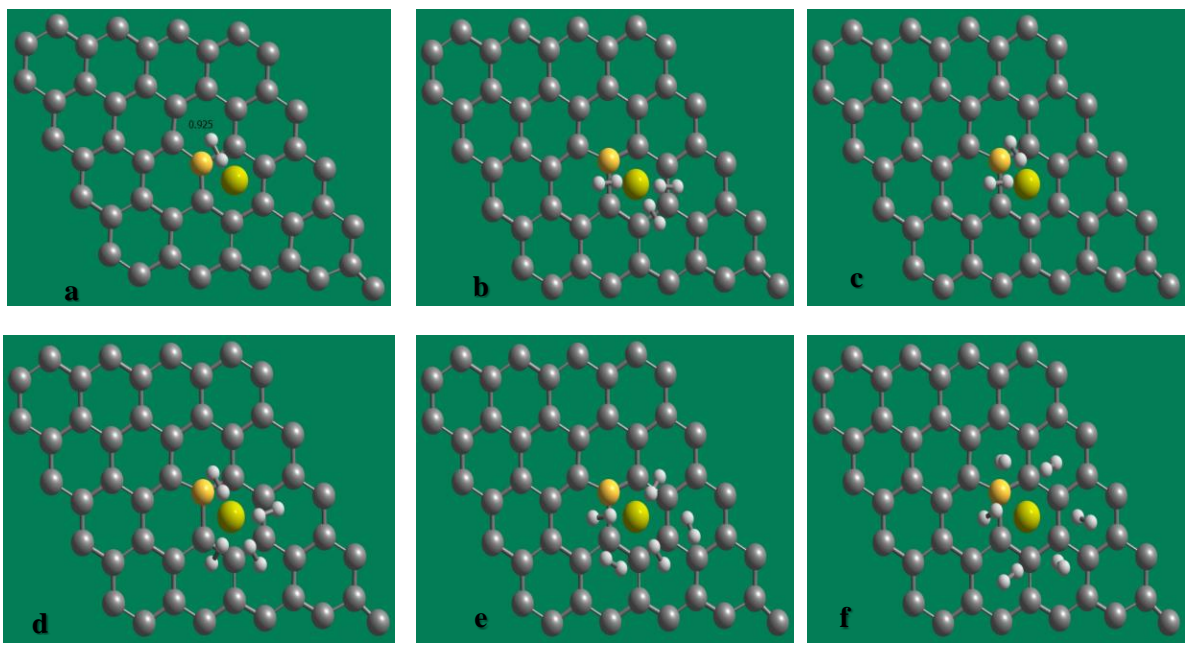

Fig.3 a-f. The optimized structures of H$_2$ molecules adsorbed on Mg/B/Graphene top view-I

The side view-II of figure 4a-f implies, there is slight distortions in all configurations particularly, in plane distortion to the graphene sheet. The lattice distortion in graphene layer [35-36] by means of adsorption of 3d-trasition metals confirms the reported slight distortions in Graphene sheet.

We know the stability H$_2$ molecules adsorbed on modified graphene sheet; we evaluated average adsorption energy of H$_2$ molecules using formula 3.

$$E_{ad-H2} = (E_{total} - E_{Mg/B/Gr} - nE_{H2})/n \qquad (3)$$

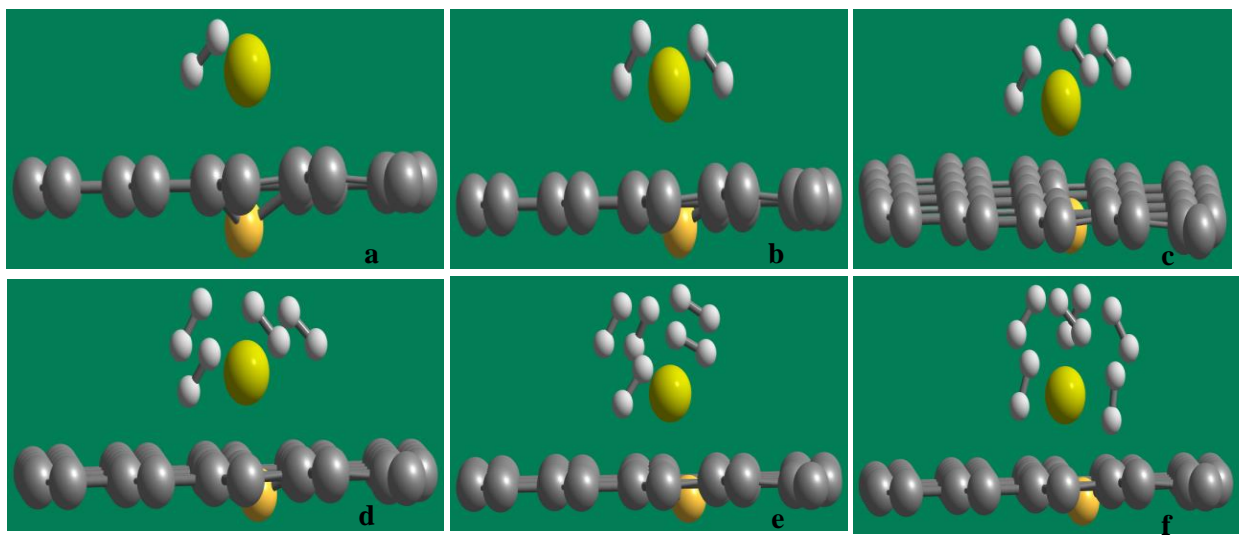

Fig.4 a-f. The optimized structures of H$_2$ molecules adsorbed on Mg/B/Graphene side view-II

| Serial Number | Number of $H_2$ molecules | Bond length of Mg/ Gr in (Å) |
|---|---|---|
| 1 | $1H_2$ | 1.949 |
| 2 | $2H_2$ | 2.296 |
| 3 | $3H_2$ | 2.412 |
| 4 | $4H_2$ | 2.225 |
| 5 | $5H_2$ | 1.996 |
| 6 | $6H_2$ | 2.249 |

Table 1 optimized bond lengths of the Magnesium decorated graphene (Mg/Gr) sheet for one to six adsorbed hydrogen molecules per Mg atom.

The extracted values of bond lengths in Å of graphene sheet for one to six hydrogen molecules per Mg atom are tabulated in table 1.The The extracted data shows there is continuous increase in bond length, when hydrogen molecules adsorbed on Mg decorated Gr sheet upto three hydrogen molecules afterwards slightly smaller in the fourth configuration (2.225Å in $4H_2$) ,fifth configuration (1.996 Å in $5H_2$),However, we increased adsorbed number of hydrogen molecules on surface of Mg decorated Gr sheet, But slightly increase in bond length is observed at sixth configuration (2.249 Å in $6H_2$).

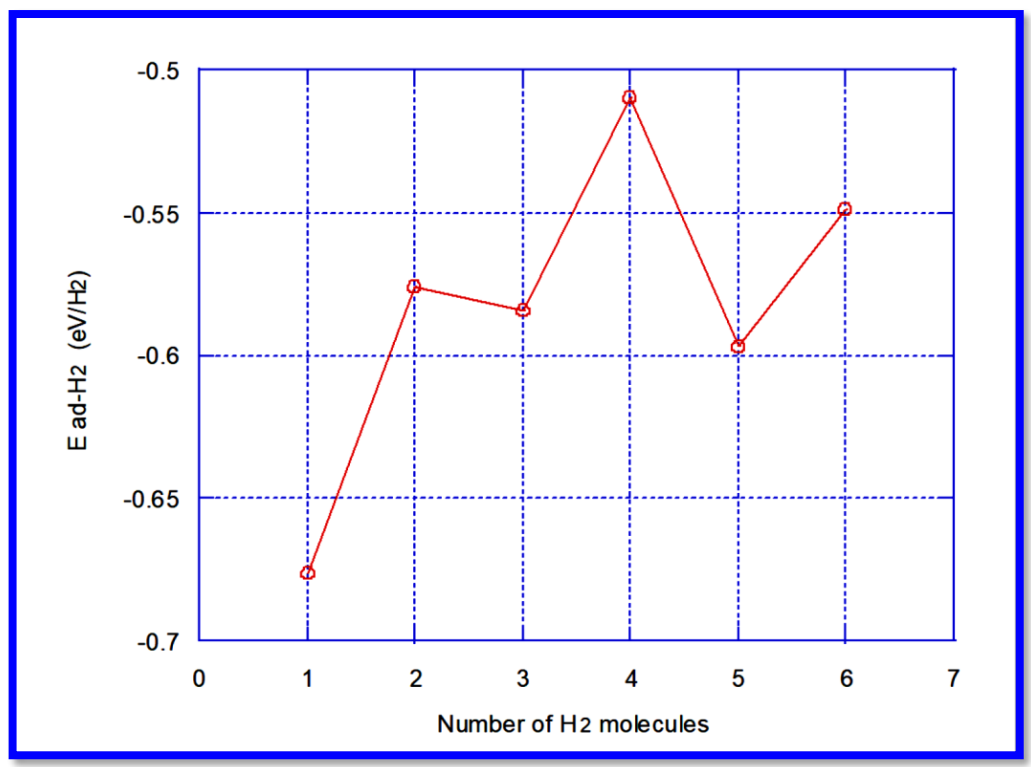

**Figure 5**. The average adsorption energies of hydrogen molecules on graphene sheet

The average adsorption energies of hydrogen molecules on graphene sheet shown in fig.5. from the plot highest value of adsorption energies is observed at $4H_2$ molecules.

## Properties of Mg on hydrogen storage

To investigate the phenomenon of magnesium atom influence, on hydrogen storage mechanism of proposed study. we characterize interactions among Mg atom with graphene and hydrogen ($H_2$) molecules. Our results imply that, the boron atom accept charge from carbon atoms and increases adsorption energy ($E_{ad}$) of magnesium atom on Gr sheet due to presence of available empty p orbitals. However, the Mg atom can adsorbed stably as well. We investigated partial density of states of complex system $1H_2/Mg/B/Gr$ as depicted in figure 5.

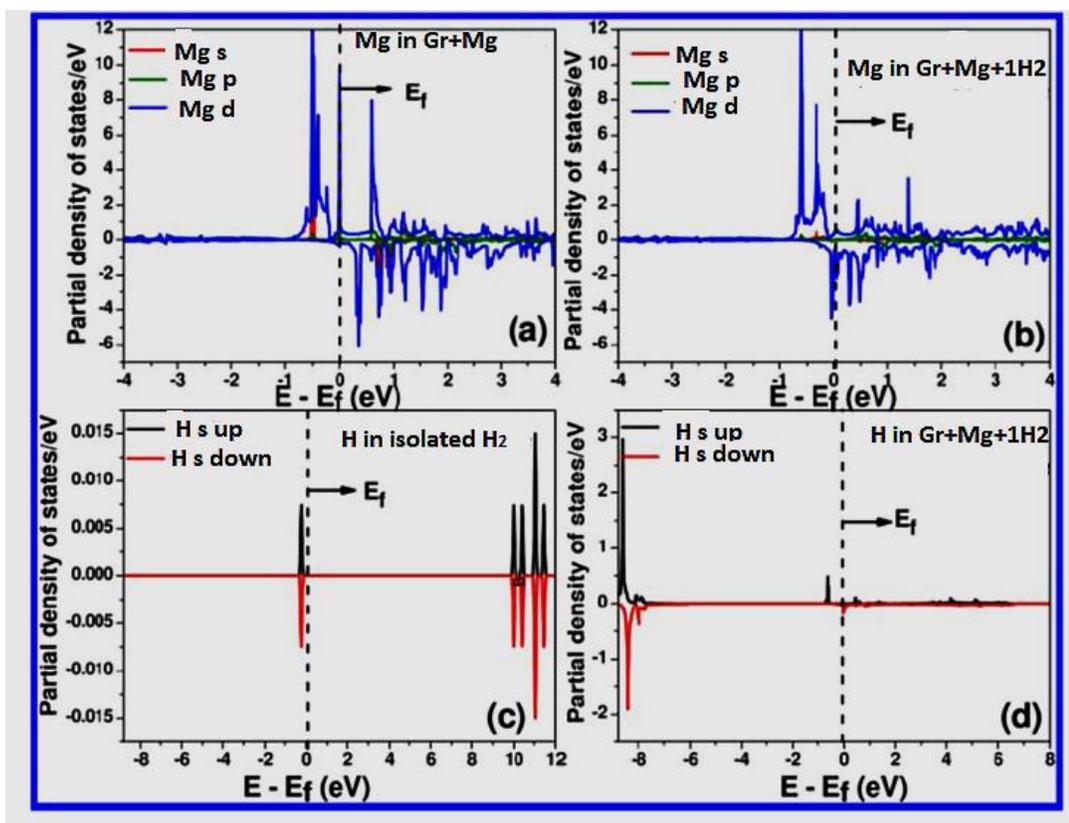

**Figure 6**. Partial density of states (PDOS) $1H_2$/Mg/B/Gr complex system.

It is observed that, s and d orbitals of $H_2$ molecule and Mg atom at -0.1eV depicts overlaps of main peaks. It indicates that strong hybridizing and binding of s and d orbitals of $H_2$ and Mg atom taking place.

There is a small amount of charge transfer the σ bonding of hydrogen molecule to d orbitals of magnesium atom. Simply, we studied the phenomenon of Mg influence atom on the process of hydrogen storage. Computational analysis performed, to reveal the interaction between Mg atom with Gr and $H_2$ molecules. The Boron (B) atom can accept charge from carbon (C) atoms and enhance adsorption energy ($E_{ad}$) of Mg atom on the surface of graphene sheet due to its empty p orbitals. However, the Mg atom can adsorb stably as well.

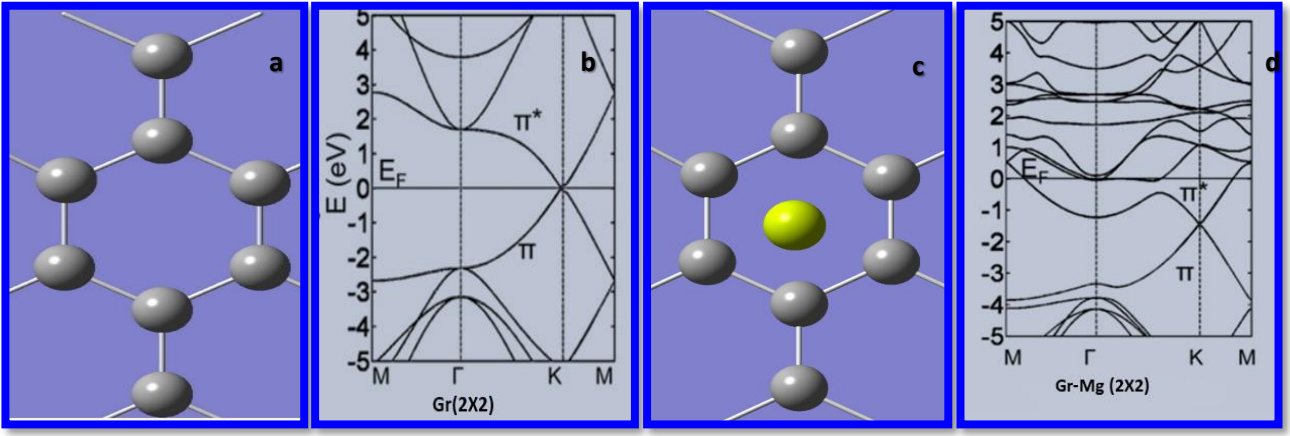

Figure 7. Gr (2X2) and Mg decorated Gr (2X2) with periodic boundary condition.

The above figure 7 shows denser coverage in Gr (2X2) and Mg decorated Gr (2X2) with periodic boundary condition.

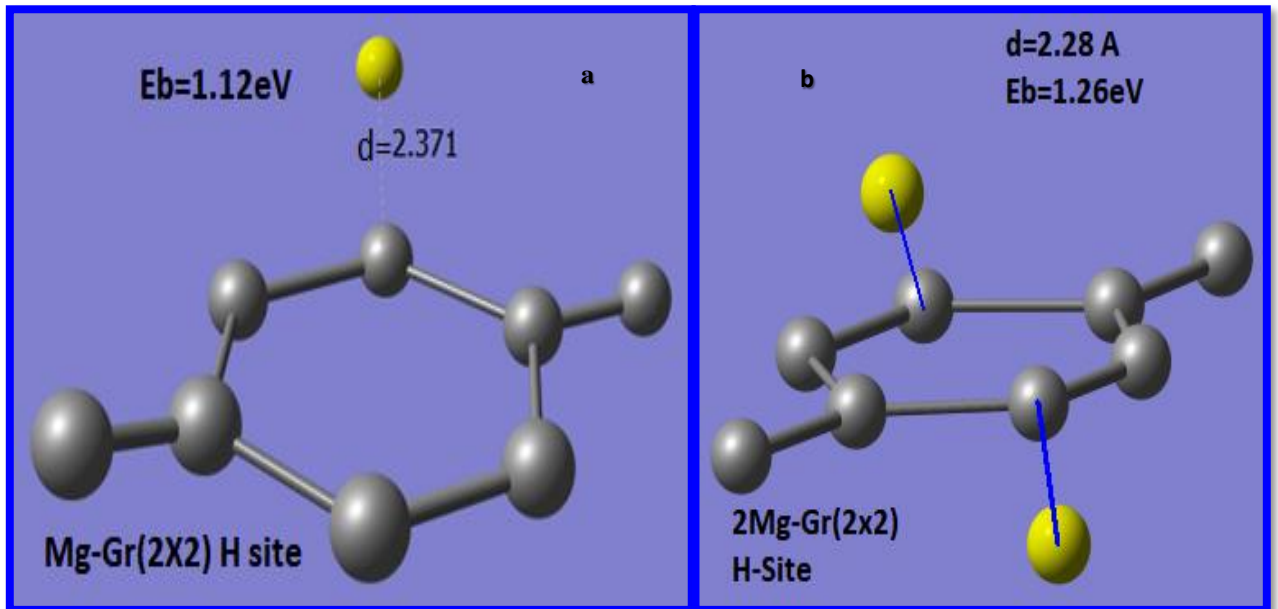

Fig.8. The optimized structure of single Mg atom (a) and double Mg atom (b) adsorbed on the H site of the (2X2) cell of graphene Mg -Gr(2X2)

Figure 8 indicates the optimized structure of single Mg atom (a) and double Mg atom (b) adsorbed on the H site of the (2X2) cell of graphene Mg -Gr(2X2).

If Mg adsorbed on one side the adsorption energy found 1.12 eV, however the Mg atom adsorbed on both side it is 1.26 eV, it is increased.

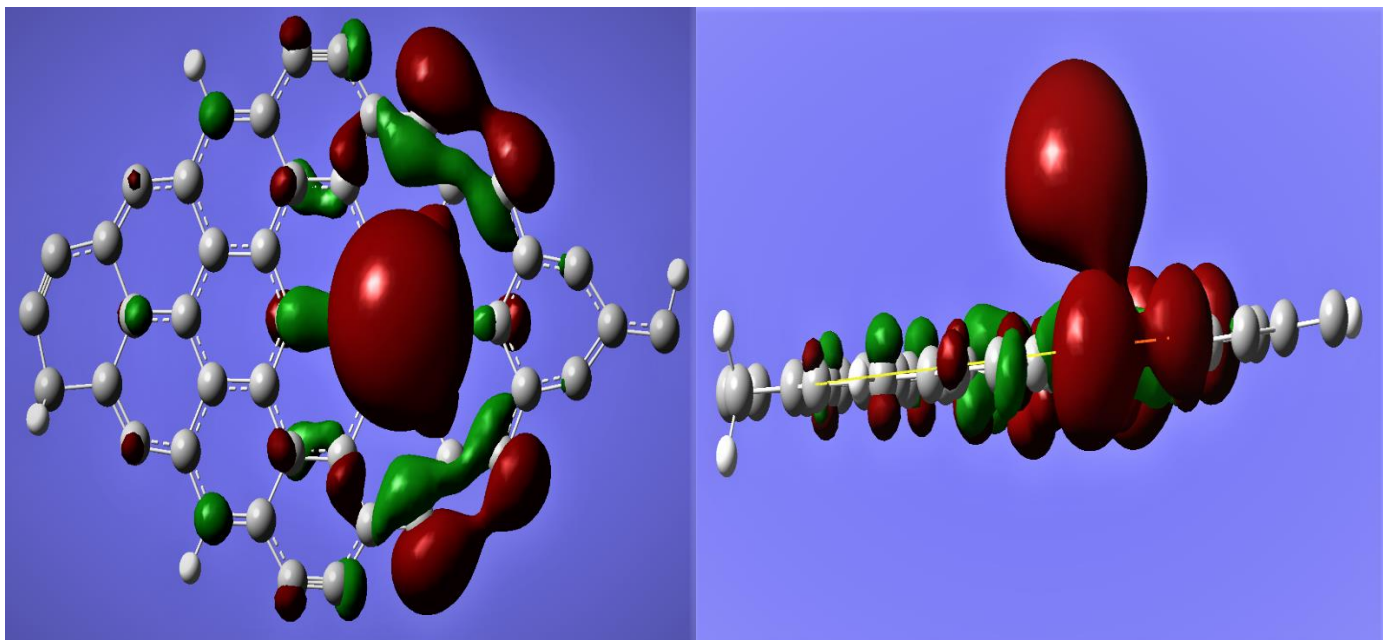

Figure 9 : Gr (4X4) B doped Mg coated complex system (a)HOMO Front view & (b)HOMO side view at MO=155;ISOVALU=0.02

Figure 9 shows (a) HOMO front view & (b) HOMO side view of the graphene (4X4) with boron doped Mg coated complex system.

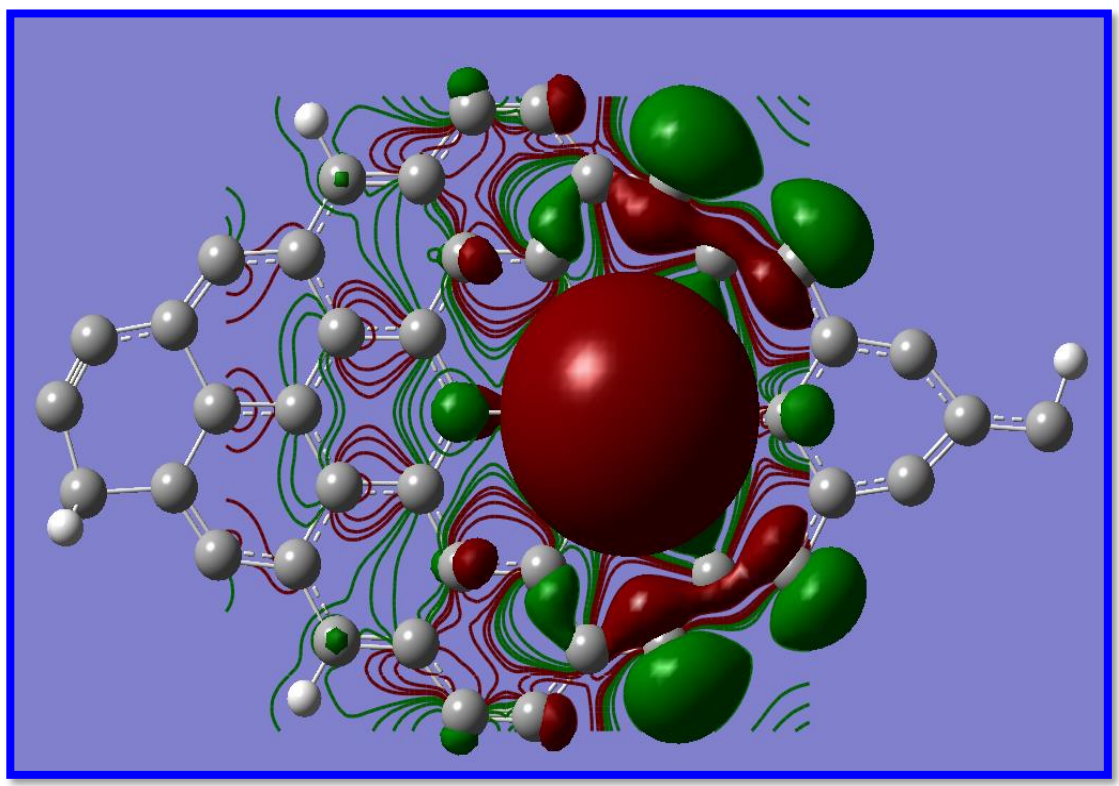

Figure 10: Gr(4X4) with Mg doped atom -LUMO MO=156  ISOVAL=0.02

Figure 10 shows the lowest unoccupied molecular orbits (LUMO) for complex system of Gr(4X4) with Mg doped atom.

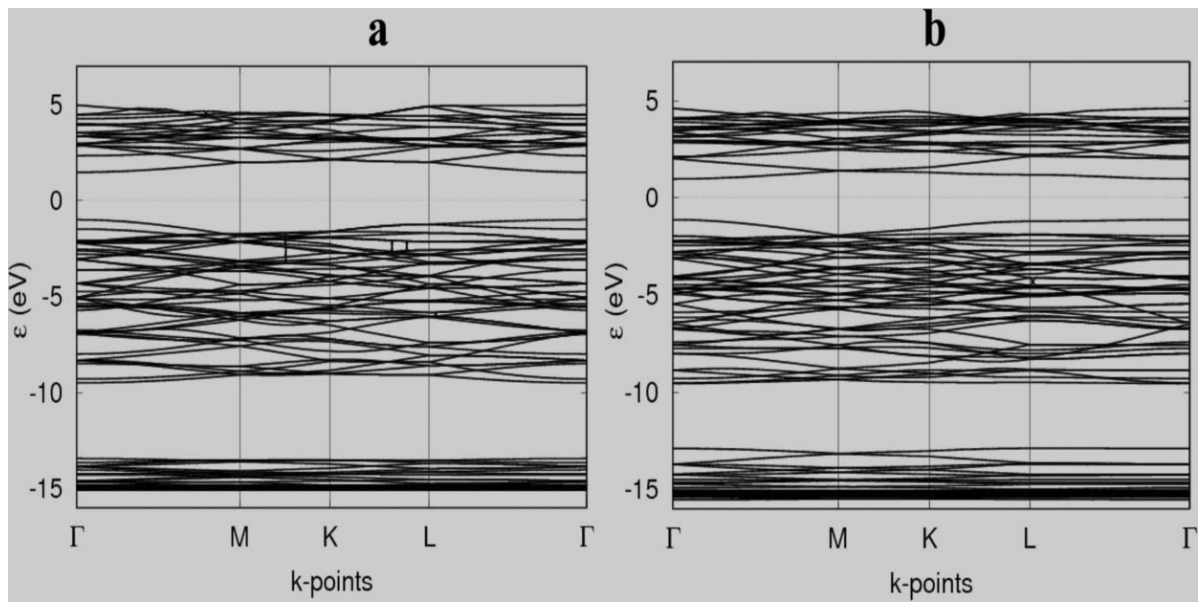

**Figure 11.** The calculated band structures at high symmetry k-points for complex system of B doped mg decorated graphene sheet. Gr-B-Mg configuration 1 (a) , Gr-B-Mg configuration 1 (b) .

The figure 11 indicates band structures at high symmetry k-points for complex system of B doped mg decorated graphene sheet. Gr-B-Mg configuration 1 (a) Gr-B-Mg configuration 1 (b).
In Fig.11 a,b. Gr-B-Mg configuration 1 and 2 reflects interaction taking between boron doped Gr by Mg atom decoration.

## 4. Conclusions

The quantum chemical analysis performed by using DFT within first principles method applied to complex system for hydrogen storage application. Our investigation suggests that Boron doped graphene with magnesium decoration achieved a uniform and stable decoration of individual B atoms on graphene sheet by Mg decoration. Our simulated data shows, there is 6H$_2$ molecules modified system can absorb with adsorption energy (E$_{ad}$) in the range between -0.510 to 0.676 eV/H$_2$. In the complex system, magnesium atom acts bridge linking H$_2$ and Gr, due to which improves adsorption capacity of Gr sheet for H storage. In complex system, we achieved 8.26 wt% gravimetric capacity of double-sided Mg decorated single B doped Gr sheet. Our findings enable excellent 2D Nanomaterial for hydrogen storage with vast outstanding potential.

## ACKNOWLEDGEMENTS

The authors are grateful to the financial support from department of science and technology (DST), New Delhi, India, under FAST TRACK SCHEME for YOUNG SCIENTIST, GRANT No. **SR/FT/LS-020/2009**(OYS 2009). We Acknowledges departmental personnel from, Nanomaterials Research Laboratory at department of Physics, Vinayakrao Patil Mahavidyalaya, Vaijapur, Maharashtra., India, for their support in using their resources.